\documentclass[11pt]{article}
\usepackage{amsmath, hyperref, wasysym}
\usepackage{latexsym}
\usepackage{cite}
\usepackage{amssymb}
\usepackage{amsthm}
\usepackage{lineno}
\usepackage[margin=0.9in]{geometry}
\usepackage{fancyhdr}
\usepackage{graphicx}
\usepackage{float}
\usepackage{color}
\usepackage{multicol}
\usepackage{authblk}
\usepackage{dsfont}
\usepackage{caption}
\usepackage[T1]{fontenc}
\usepackage{lmodern}
\usepackage{booktabs}
\restylefloat{table}
\begin{document}
\title{Methods for quantifying self-organization in biology: a forward-looking survey and tutorial}
\author[*]{Alexandria Volkening}
\affil[*]{Department of Mathematics, Purdue University, West Lafayette, IN, USA (\href{mailto:avolkening@purdue.edu}{avolkening@purdue.edu})}
\maketitle

\begin{abstract}From flocking birds to schooling fish, organisms interact to form collective dynamics across the natural world. Self-organization is present at smaller scales as well: cells interact and move during development to produce patterns in fish skin, and wound healing relies on cell migration. Across these examples, scientists are interested in shedding light on the individual behaviors informing spatial group dynamics and in predicting the patterns that will emerge under altered agent interactions. One challenge to these goals is that images of self-organization---whether empirical or generated by models---are qualitative. 
To get around this, there are many methods for transforming qualitative pattern data into quantitative information. In this tutorial chapter, I survey some methods for quantifying self-organization, including order parameters, pair correlation functions, and techniques from topological data analysis. I also discuss some places that I see as especially promising for quantitative data, modeling, and data-driven approaches to continue meeting in the future. \end{abstract}

\section{Introduction}
\label{sec:intro}

\begin{figure}[t!]
\includegraphics[width=0.7\textwidth]{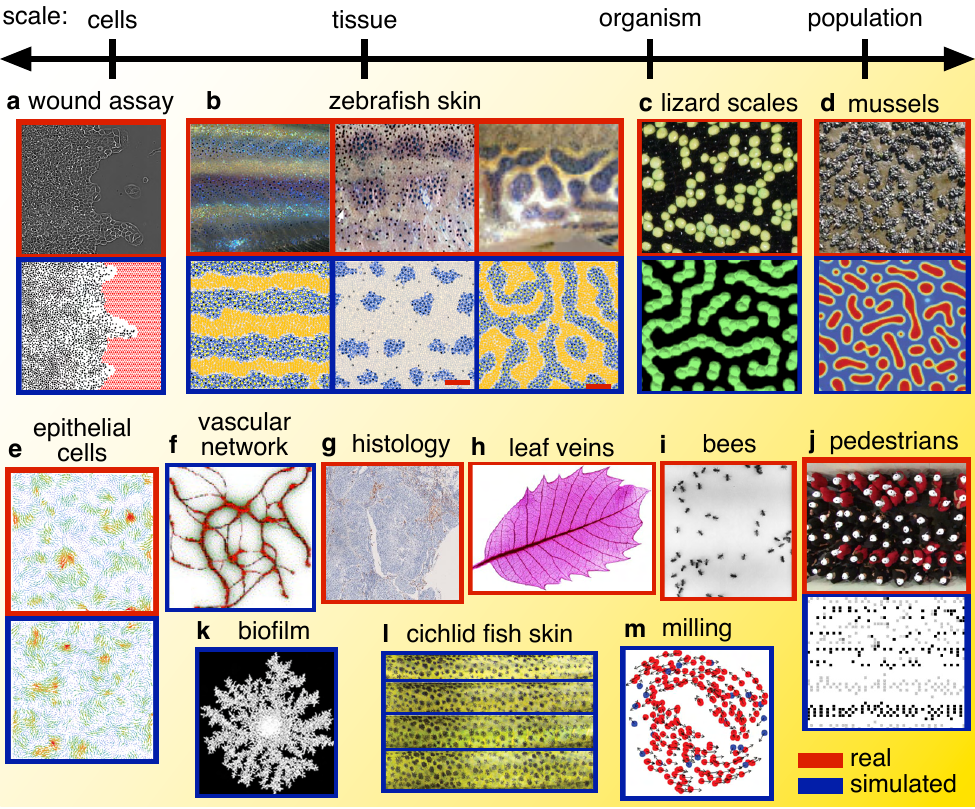}\centering
\caption{\label{fig:overview}Empirical and model-generated images of self-organization across scales. (a) In a scratch-assay experiment, cells spread into empty space \cite{Sepulveda}; (b) zebrafish feature a wide range of skin patterns, all made up of brightly colored pigment cells \cite{Frohnhofer,volkening2}; (c) lizard scales display labyrinth-style patterns \cite{Manukyan2017}; and (d) mussels organize into groups \cite{Liu2013mussel}. (e) Epithelial cells interact and align \cite{Sepulveda}, and (f) developing blood vessels exhibit branching \cite{MerksImage}. (g) Histology images indicate cells in tissue \cite{Bull2020}. (h) Veins form branching patterns \cite{MerksImage}, and (i) honeybees interact and exchange food \cite{PelegTDA2023}. (j) Lanes may emerge when pedestrians move in opposite directions \cite{Sieben2017,Zhang2012}. (k) Biofilms exhibit branching \cite{Biofilm2023}, and (l) cichlid fish feature messy stripes, which form during development \cite{Kratochwil}. (m) Flocking, swarming, and milling occur in many biological systems \cite{Bhaskar2019,Couzin2003,VicsekReview,Dorsogna2006}. Images in (a, e) reproduced from \cite{Sepulveda}, Copyright Sep\'{u}lveda et al., and licensed under CC BY; top row of (b) reproduced from \cite{Frohnhofer}, published by The Company of Biologists Ltd.; and licensed under CC-BY $3.0$ (\url{https://creativecommons.org/licenses/by/4.0/}); bottom row of (b) reproduced from \cite{volkening2} and published by Springer Nature; images in (c) reproduced from \cite{Milinkovitch2021}; images in (d) reproduced from \cite{Liu2013mussel} with permission from the Proceedings of the National Academy of Sciences, U.S.A.; image in (f) reproduced from \cite{MerksImage} with permission of Elsevier, Copyright (2006) Elsevier Inc.; image in (g) reproduced from \cite{Bull2020}; image in (h) reproduced from \cite{Iwamasa2023}; image (i) reproduced from \cite{PelegTDA2023}; top panel of (j) reproduced from \cite{Sieben2017}; bottom panel of (j) reproduced from \cite{Burstedde2001} with permission from Elsevier, Copyright (2001) Elsevier Inc.; image in (k) reproduced from \cite{Biofilm2023} with permission of Elsevier, Copyright (2023) Elsevier Inc.; images in (l) reproduced (cropped) from \cite{Kratochwil}, Copyright Liang et al.; image in (m) reproduced from \cite{Bhaskar2019}. Images in (c, g, h, i, l, m), the bottom row of (b), and the top panel of (j) licensed under CC-BY $4.0$ (\url{https://creativecommons.org/licenses/by/4.0/})}
\end{figure}

Pattern formation driven by the interactions of agents is found across the natural and social world, spanning the population scale to the intracellular scale. Large-scale examples include pedestrian movement in crowds \cite{Sieben2017,Zhang2012,Helbing1995,Burstedde2001}, honeybee aggregation \cite{PelegTDA2023}, schooling fish \cite{Katz2011fish,VicsekReview}, and marching locusts \cite{Buhl2006}. On the scale of cells and tissues \cite{Buttenschon2020}, neural-crest cell migration \cite{Giniunaite2020}, color pattern formation in fish skin \cite{Nakamasu,VolkeningRev}, and cell dynamics during wound healing are examples of self-organization. At still smaller scales, proteins and filaments regulate transport and shape within cells \cite{Ciocanel2021}. Studying such complex systems---whether empirically or through mathematical modeling---often leads to qualitative pattern data in the form of images. Being able to quantify these spatial data opens the door to a broader perspective on self-organization and makes complex systems more amenable to interdisciplinary investigation. With this motivation, in this tutorial chapter written for an interdisciplinary audience, I discuss a selection of methods for making qualitative data quantitative that apply to disparate applications.

Across applications in self-organization, a general goal is to describe agent interactions and predict how altered behaviors affect collective dynamics. In a developmental-biology setting, doing so could mean understanding how genes regulate cell behavior and organism phenotype. In a pedestrian-movement example, shedding light on crowd dynamics could help engineers design buildings that allow for quicker evacuation during emergencies. Studying such systems often involves developing predictive models and characterizing the group dynamics that arise from these models under different parameter values and model rules. In some cases, models that do not describe space---such as compartmental models that track the number of agents in time---are appropriate, and, in other settings, the problem calls for models that treat agent behavior in time and space. In this chapter, my focus is on quantifying spatial patterns, so I consider the latter only. As I show in Fig.~\ref{fig:overview}, the task of quantifying qualitative pattern data is largely the same regardless of whether the data are collected empirically or generated by a spatial model. The main difference is that working with empirical images involves the additional step of first identifying the agents of interest (i.e., pedestrians in Fig.~\ref{fig:overview}(j)) in the data, whereas models provide full information on agent positions or density; also see Fig.~\ref{fig:data}.

Quantifying pattern data serves many goals in interdisciplinary studies. For example, when data and a model are available, one may be interested in estimating model parameters \cite{Toni2009ABC,Bode2019}. Or the goal may be to use data-driven approaches (i.e., \cite{Mangan2017,Brunton2016,Nardini2021,Kemeth2022}) to select model terms from a large library of possible terms, based on some ground-truth data. There are also cases where researchers seek to develop algorithms for producing realistic synthetic data in order to train segmentation algorithms. Across these examples, there is a need to measure the distance between empirical data and model-generated output, necessitating some quantitative description of the empirical and \textit{in silico} patterns \cite{Pargett2013}. In other cases, the data may be fully synthetic or fully empirical. In medical settings, distinguishing between images of healthy and unhealthy tissue is a natural goal. Or, in the case of modeling data, one may want to comprehensively characterize the patterns that can emerge from a model under changes to its parameters. In still other settings, the goal may be to understand the relationship between different modeling approaches to the same problem in order to determine how implementation choices affect predictions \cite{Plank2012,Osborne2017,Cleveland,Kursawe2017}.

All of these goals boil down to a need for transforming data from qualitative to quantitative and quantifying the difference between patterns, whether synthetic and empirical, both empirical, or both synthetic. This a fundamental challenge that arises across disciplines, including biology, ecology, mathematics, engineering, and social science. There are thus many, many approaches to quantifying spatial patterns, and specific methods are more or less common in each research community. This motivates my chapter. By discussing shared features in pattern data and introducing a range of quantification methods, 
my goal is to support cross-fertilization between applications and perspectives. First, in \S\ref{sec:data}, I discuss how to express pattern data from various sources---whether empirical images, agent-based models, or continuum models---as point clouds. I follow this in \S\ref{sec:quantify} by highlighting some quantifiable patterns features that are present across applications. I next overview a selection of quantification methods, covering order parameters (\S\ref{sec:order}), pattern-simplicity scores (\S\ref{sec:pair}), pair-correlation functions (\S\ref{sec:pair}), and techniques from topological data analysis (\S\ref{sec:tda}). I conclude in \S\ref{sec:discussion} with some comments on data-driven modeling.

In terms of the themes of this book, which includes the mathematics of animal ecology and cell biology, I come from the cell-biology perspective. I stress that this chapter is a selection of some of the quantification methods that I have encountered as a mathematical modeler, mainly of patterning in developmental biology. Importantly, there are a wealth of methods for quantifying spatial data that I do not discuss, such as many in spatial statistics, network-based analysis, Fourier analysis, and morphometric tools. For additional references, I highlight \cite{Feng2023,Bull2020,Law2009,Bull2024preprint,VicsekReview,morph,Summers2022}. For the approaches that I discuss, I seek to provide a starting point, suggest references to learn more, and encourage work that brings together multiple methods to quantify the same data and actively considers the choices involved. In writing this chapter, I have reflected on what the various approaches in \S\ref{sec:methods} could say about my pattern data. I hope this chapter and the references herein encourages readers to think about their data from a new perspective as well.

\section{Obtaining point-cloud data on self-organization}\label{sec:data}

\begin{figure}[t!]
\includegraphics[width=0.7\textwidth]{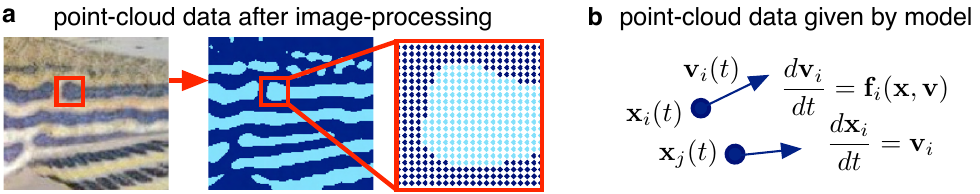}\centering
\caption{\label{fig:data} Point-cloud data in self-organization. (a) One can use image-processing or machine-learning approaches to process empirical image data. (I used the free software Ilastik \cite{Ilastik} to identify dark and light regions of this fish skin pattern.) After taking the centers of the pixels of interest in the processed image, the result is point-cloud data in the form of $(x,y)$-coordinates for each light-blue and navy pixel. A similar approach can be used with patterns generated by continuum models. (b) In the case of agent-based models, point-cloud data are often immediately available. Here the position $\textbf{x}_i(t)$ and velocity $\textbf{v}_i(t)$ of the $i$th agent are specified by the model itself. Left image in (a) reproduced from \cite{Frohnhofer}, published by The Company of Biologists Ltd.; and licensed under CC-BY $3.0$ (\url{https://creativecommons.org/licenses/by/4.0/}); I added the red box and cropped the image.}
\end{figure}

Throughout this chapter, I mainly discuss quantifying pattern data that are in the form of point clouds, meaning collections of points, typically in two dimensions. As I show in Fig.~\ref{fig:data}(a), one can arrive at point-cloud data from an empirical image by thresholding, applying other image-processing tools, or using machine-learning methods; in this example, I used the free software Ilastik \cite{Ilastik} to hand-select a few regions of the empirical image in Fig.~\ref{fig:data}(a, left panel), and then the software classified the pixels in the rest of the image automatically. After limiting my focus to the pixels with a given color value, the result is point-cloud data on a square grid. Specifically, in Fig.~\ref{fig:data}(a, right panel), I plot light-blue and navy points indicating the $(x,y)$-coordinates of the centers of each light-blue or navy pixel.

For patterns generated by simulating mathematical models, the approach to arriving at point-cloud data depends on the framework. There are two particularly common frameworks for modeling self-organization spatially, namely agent-based (or individual-based) and continuum models \cite{VolkeningRev}. Continuum models track agent density; in spatial settings, these take the form of partial differential equations or integro-differential equations. Reaction-diffusion equations (i.e., \cite{Gaffney2019,Turing,Mein,Menshykau2019}), Cahn--Hilliard equations (i.e., \cite{Liu2013mussel,Cahn}), and non-local partial differential or integro-differential equations (i.e., \cite{Carrillo, painter, Bernoff2011,Mogilner1999}) have all been used to describe the evolution of agent density. Partial differential equations are often studied through bifurcation analysis, and researchers can quantitatively describe the most unstable wavelengths. In terms of the data that they produce, continuum models of patterning generally provide the density $\rho(\textbf{x},t)$ (i.e., of agents) in space $\textbf{x}$ and time $t$; I show an example of these data in the bottom panel of Fig.~\ref{fig:overview}(d). In order to produce point-cloud data from simulations of continuum models, the approach is similar to working with empirical images---thresholding a continuous agent density leads to point-cloud data.

Off-lattice agent-based (i.e., \cite{volkening, volkening2}) and on-lattice agent-based models (i.e., \cite{Bullara, Owen2020}) take a more microscopic perspective, describing the interactions of individual agents during self-organization and offering detailed predictions; see the bottom panels of Fig.~\ref{fig:overview}(a) and Fig.~\ref{fig:overview}(b) respectively, for examples of patterns generated by on- and off-lattice agent-based models. Agent-based models are common in animal-ecology studies of flocking, swarming, and schooling, as well in research on the behavior of cells in developing and regenerating tissues. Stochastic agent-based models tend to have many parameters, and it is very difficult to broadly characterize the full range of patterns that these models can produce. As I show in Fig.~\ref{fig:data}(b), the data generated by simulating off-lattice agent-based models (sometimes called ``interacting particle systems") is typically already in point-cloud form as $\{\textbf{x}_i(t)\}_{i=1,2,...,N}$, where $\textbf{x}_i(t)$ is the position (e.g., $(x,y)$-coordinates) of the $i$th agent at time $t$. In other cases, we also have the velocity $\textbf{v}_i(t)$ of each agent. For on-lattice agent-based models, space is discretized, and the pattern data may be lists of grid positions occupied by specific agents at each time; one can use the centers of each grid position of interest to build a point cloud. Other types of microscopic models, such as vertex-based models and cellular Potts models \cite{Glazier1993,Merks2017}, can also give rise to point-cloud data.

\section{Shared features in biological patterns}\label{sec:quantify}

\begin{figure}[t!]
\includegraphics[width=0.7\textwidth]{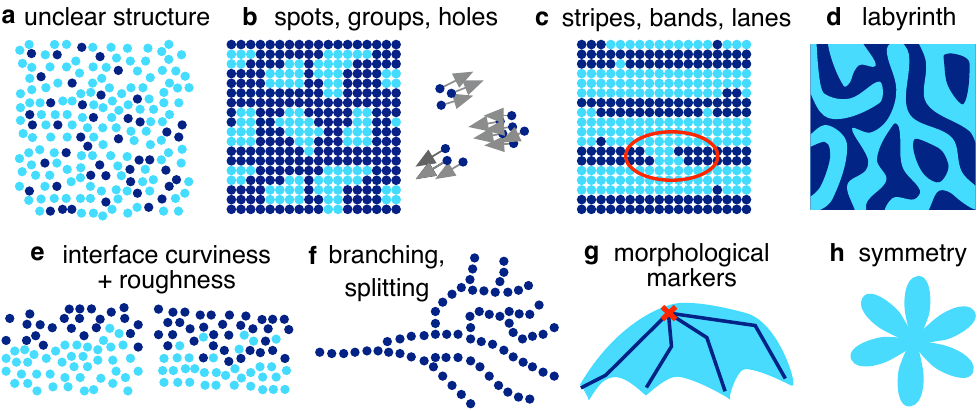}\centering
\caption{\label{fig:summary}Some example features in qualitative data that researchers often seek to quantify. (a) In some cases, the structure in pattern data is difficult to describe or qualitatively judge. Here it is not clear whether the navy points are distributed uniformly at random or if additional structure is present. (b) Spots, clusters, and groups appear in many of the patterns in Fig.~\ref{fig:overview}, as do (c) stripes, bands, and lanes. In (c), I highlight that identifying interruptions in stripes can be important in some contexts. (d) Labyrinth patterns may feature a characteristic wavelength without directionality. (e) Whether working with the wound assay in Fig.~\ref{fig:overview}(a) or stripes in fish skin as in Fig.~\ref{fig:overview}(l), researchers may quantify curviness or roughness of stripes, interfaces, and boundaries. (f) The vascular network and leaf veins in Fig.~\ref{fig:overview} are examples of branching, bifurcating, or splitting patterns. One may also view the labyrinth pattern in Fig.~\ref{fig:overview}(b) as branching, and, in some sense, the navy points in panel (c) of this figure are a branching pattern. (d) In a different vein (no pun intended), the position of morphological markers are important to describe in some biological images \cite{morph}. (e) Symmetry (of various kinds) is the feature of interest in other examples of self-organization \cite{Gandhi2021}.}
\end{figure}

Before discussing similarities in patterns across scales in biology, I find it useful to highlight three broad perspectives on quantifying patterns, though the distinctions are not completely clearcut. First, in some cases, scientists study patterns without obtaining interpretable quantitative descriptors. For example, machine-learning methods can be used to identify features of clinical images that are related to health or disease \cite{Bull2024}. Second, there are patterns in which the specific positions of points in space matter, and then the difference in density in space may be a suitable measurement to compare patterns. For example, in some scratch-assay experiments, it is sufficient to consider a cross-section (or mean cross-section) of the density of cells spreading into empty space \cite{Lagergren2020}. Third, in messy biological data and stochastic simulations, often it is pattern features---and not precise positions---that are important. For example, in the fish pattern in Fig.~\ref{fig:overview}(b), one may want to capture that blue spots are present and roughly aligned in stripes, but the individual position of any given spot could be shifted \cite{McGuirl2020}. In this chapter, I focus mainly on the third perspective.

In order to apply methods from disparate applications to one's data, it is useful to look for common features in Fig.~\ref{fig:overview}. I summarize some of these features in Fig.~\ref{fig:summary}, and note that their names differ depending on the application and field. For example, clusters of points like those in Fig.~\ref{fig:overview}(b, center), Fig.~\ref{fig:overview}(d), or Fig.~\ref{fig:overview}(e) may be called connected components, spots, or groups. Stripes, lanes, trails, or bands are present in many of the patterns in Fig.~\ref{fig:overview}, sometimes with sharp interfaces, and in other cases (i.e., Fig~\ref{fig:overview}(l)) with rougher, less clear boundaries \cite{Kratochwil}. Branching, bifurcating, and splitting is another common feature of biological patterns, including leaf venation \cite{Iwamasa2023,Rolland2009}, blood-vessel networks \cite{MerksImage}, and yeast colonies \cite{Binder2015yeast}. From a different perspective, networks of blood vessels can also be described as labyrinthine, as can the fish-skin patterns in the right panel of Fig.~\ref{fig:overview}(b). Various types of symmetry \cite{Gandhi2021} are widespread in natural patterns as well. Finally, quantifying biological patterns and image data involves landmarks or morphological markers in some cases. 

\section{A selection of methods for quantifying patterns}\label{sec:methods}

I now turn to a selection of methods for quantifying pattern data like those in Fig.~\ref{fig:summary}(a)--(f). Quantifying morphology and symmetry as in Fig.~\ref{fig:summary}(g)--(h) are outside of the scope of this chapter; I highlight reference \cite{Vasyl2021} as an example of landmark-free analysis of morphology and the work of Gandhi et al. \cite{Gandhi2021} for more information about symmetry in pattern data. I introduce order parameters in \S\ref{sec:order}, pattern simplicity scores in \S\ref{sec:simplicity}, pair-correlation functions in \S\ref{sec:pair}, and topological techniques in \S\ref{sec:tda}. Except for some examples in \S\ref{sec:simplicity}, I consider pattern data in the form of point clouds, typically in two dimensions. As I discuss in \S\ref{sec:intro}, the approaches that I illustrate next are a subset of the many quantification methods that researchers use across disciplines.

\subsection{Order parameters for point-cloud data}
\label{sec:order}

Order parameters have been used to quantify point-cloud data in studies of aggregation \cite{VicsekReview,Topaz2015} across the spectrum of biological scales, including intracellular dynamics \cite{Ciocanel2021}, cell interactions \cite{Bonilla2020}, and animal or insect populations \cite{Couzin2002,Buhl2006}. Broadly, order parameters provide a summarized statement of the degree of order in a system and its overall dynamics; they are particularly useful in identifying transitions from one state or phase to another \cite{Order,VicsekReview}. What order parameter is most appropriate to describe any given pattern data is application dependent. As Topaz et al. \cite{Topaz2015} note, this means order parameters are often defined on a case by case basis, using information about the global dynamics that one seeks to identify in data. (Other quantification approaches, like those I discuss in \S\ref{sec:pair} and \S\ref{sec:tda}, apply more broadly, though there are also choices involved in these methods.)

In many cases, the point-cloud data that researchers are interested in quantifying using order parameters are in the form of agent positions and/or velocities, as in Fig.~\ref{fig:overview}(m) and Fig.~\ref{fig:data}(b).) One common order parameter for studies of aggregation introduced by Vicsek et al.\ \cite{Vicsek1995} is the normalized average velocity across the members of the group, as below:
\begin{align}
\phi(t) &= \text{normalized average velocity} = \frac{1}{Nv_\text{avg}} |\sum_{i=1}^N \textbf{v}_i(t)|,\label{eq:phi}
\end{align}
where $N$ is the number of group members, $v_\text{avg}$ is their average speed, and $\textbf{v}_i(t)$ is the velocity of the $i$th agent at time $t$ \cite{Topaz2015,VicsekReview,Vicsek1995}. When all members of the group are moving in nearly the same direction, the value of $\phi(t)$ is near one, and, when all of the agents are moving randomly, $\phi(t)$ is close to zero \cite{Topaz2015,VicsekReview}. Importantly, if two groups of agents are moving in opposite directions, it is also possible for the normalized average velocity to be zero \cite{Topaz2015}, indicating how this order parameter is able to capture meaningful information in some cases but not in others.

Using multiple order parameters together can provide a means around this difficulty and increase the number of patterns that one can distinguish. Other order parameters include the polarity and normalized angular momentum \cite{Chuang2007,Couzin2002}, and Chuang et al. \cite{Chuang2007} developed a modified normalized angular momentum order parameter that is able to capture double mills. Together all of these order parameters provide information about coherently moving flocks, single mills, and double mills \cite{Chuang2007}. As in the example in Eqn.~\eqref{eq:phi} for pattern data in an ordered phase, the value of its associated order parameter is typically non-zero (and often bounded by one), whereas order parameters are generally defined so that they are zero when disorder is present \cite{VicsekReview}. Order parameters are not always bounded between zero and one, however, and quantities such as the mean distance between nearest neighbors in aggregation data can also be considered an order parameter \cite{Huepe2008}.

\subsection{Pattern simplicity scores and measurements for segmented data}
\label{sec:simplicity}

For qualitative data in which the pattern features have been segmented and identified, there are a wealth of summary statistics that one can compute. In the case of Fig.~\ref{fig:summary}(b, left panel), for example, manual labelling, machine-learning approaches, image-processing tools, clustering methods, or topological techniques \cite{McGuirl2020} (see \S\ref{sec:tda}) may be used to identify twelve spots, each made up of a set of light-blue pixels. One can then count the number of points per spot and calculate the average distance between spot centers. Similarly, in Fig.~\ref{fig:summary}(c), once the navy pixels are each assigned to a stripe feature, methods from computational geometry \cite{Esurvey} (as in \cite{Cleveland}) may be used to quantify stripe curviness. For studies of branching blood-vessel networks as in Fig.~\ref{fig:summary}(f), researchers have arrived at many summary statistics, including inter-vessel spacing, the number of branches or vessel endpoints, and the length of the longest vessel \cite{Byrne2019arxiv,Nardini2021arxiv}. When working directly with image data, the built-in methods in the powerful open-source software ImageJ \cite{ImageJ} and Fiji \cite{Fiji} offer a wide range of quantitative descriptors for segmented image features.

Because they lend themselves to distinguishing between spot and labyrinth patterns, I highlight pattern simplicity scores \cite{Miyazawa2010Simplicity} as a useful summary statistic that can be computed given labelled data on pattern features. Specifically, in Fig.~\ref{fig:summary}(d), there are five clusters of light-blue pixels, viewed as five continuous regions of light blue. Depending on the community, these five clusters may be called ``connected components" or ``contours", and I refer to them broadly as features. Let $A_i$ be the area of the $i$th feature, and $P_i$ be its perimeter, where $i=\{1,2,3,4,5\}$ in our example since there are five light-blue features in Fig.~\ref{fig:summary}(d). Then the pattern simplicity score (PSS) is the average circularity of the features, weighted by their area, as below:
\begin{align*} \text{PSS} &= \sum_{i=1}^{\text{number of pattern features}} {\frac{A_i}{\sum_{j=1}^{\text{number of pattern features}}A_j}} ~~~\times ~~~{ \frac{4 \pi A_i}{P_i^2}}.
\end{align*}
Miyazawa et al. \cite{Miyazawa2010Simplicity} used the PSS to study different types of fish, and Djurdjevic et al. \cite{Djur2019Trout}, focused on pattern simplicity scores and overall color tone in images of trout. The researchers \cite{Miyazawa2020} considered the pattern complexity score---which is one minus the PSS---to quantify skin-pattern diversity across a very wide range of fish.

\subsection{Pair-correlation functions}
\label{sec:pair}

\begin{figure}[t!]
\includegraphics[width=0.7\textwidth]{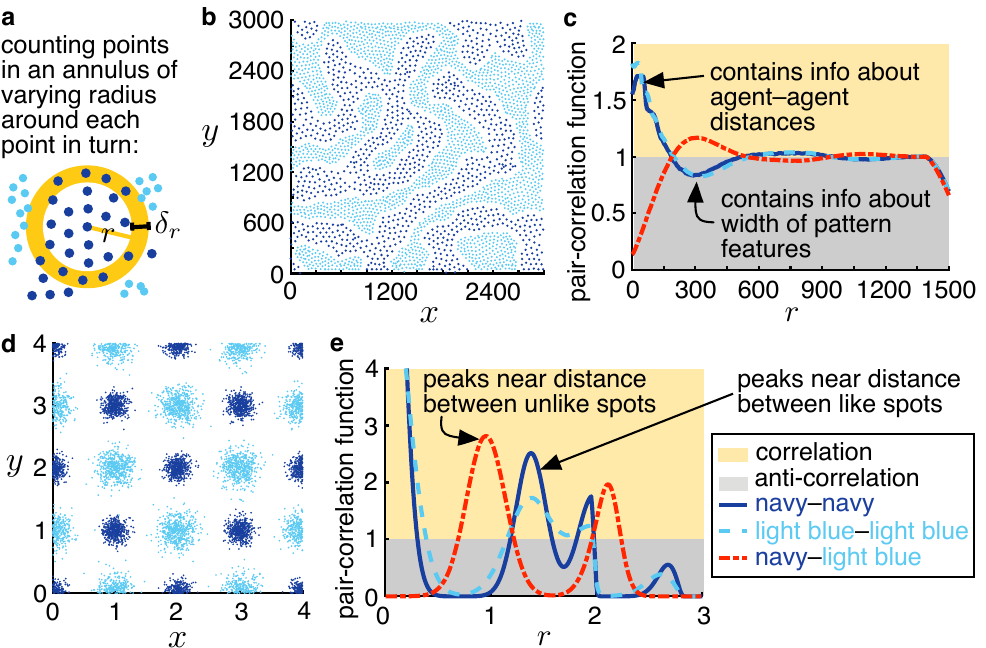}\centering
\caption{\label{fig:pcf} Overview of pair-correlation functions. (a) Applying pair-correlation functions to point-cloud data that are continuous in space broadly involves counting the number of points of different types that appear in an annulus center around each point \cite{Bull2020}. (b) I show an example labyrinthine pattern in a domain that is periodic in $x$ and $y$, and the corresponding PCFs and cross-PCF according to Eqns.~\eqref{eq:pcf1}--\eqref{eq:pcf2}. (d) As another example, I show a spotted pattern in a periodic domain, as well as (e) the associated PCF summaries.}
\end{figure}

Pair-correlation functions (or radial distribution functions) \cite{Bull2024,Bull2020,Binder2013,PCFbook,Binder2015,Dini2016,Treloar2014spatial} belong to the field of spatial statistics and can provide quantitative summaries of a wide range of patterns, including those with spots, stripes, labyrinth features, branching, and unclear structure in Fig.~\ref{fig:summary}. Broadly, pair-correlation functions (PCFs) help researchers distinguish between aggregation, segregation, and spatial randomness in point-cloud data \cite{Gavagnin2018yates}, and their output can also be further interpreted to describe more pattern features, as I show in Fig.~\ref{fig:pcf} and discuss in this subsection. There is a large literature on PCFs and their application to pattern data, and my discussion below highlights a subset of these directions.

Pair-correlation functions work by describing the density of points at different scales, relative to complete spatial randomness (that is, points appearing completely at random in the domain of interest) or some other null distribution \cite{Bull2020,Bull2024}. As I show in Fig.~\ref{fig:pcf}, the PCF approach involves counting the number of points of a given type in an annulus of inner radius $r$ and width $\delta_r$ centered at each point of that type, and the result of this process is information about clustering and characteristic length scales \cite{Agnew2014}. Similarly, cross-PCFs quantitatively summarize the relationship between points of two types (i.e., light blue and navy in Fig.~\ref{fig:pcf}) in data. Ripley's $K$ index \cite{Ripley1977} is a related measurement in spatial statistics that counts points in circles of increasing size, rather than in annuli.

The specific definitions of the PCF and cross-PCF depend on the choice of null distribution and approach to boundary conditions \cite{Bull2020}. In Fig.~\ref{fig:pcf}, I apply the PCF and cross-PCF definitions of Bull et al. \cite{Bull2020} to particles of two types, as below:
\begin{align}
\text{PCF}(r) &= \frac{A}{N_\alpha^2}\sum_{i=1}^{N_\alpha} \frac{1}{A_r(\textbf{x}^\alpha_i)}\sum_{j=1}^{N_\alpha} \mathds{1}_{\delta_r}( \| \textbf{x}^\alpha_i - \textbf{x}^\alpha_j \| -r), ~\text{and} \label{eq:pcf1}\\
\text{cross-PCF}(r) &= \frac{A}{N_\alpha}\sum_{i=1}^{N_\alpha} \frac{1}{N_\beta} \frac{1}{A_r(\textbf{x}^\alpha_i)}\sum_{j=1}^{N_\beta} \mathds{1}_{\delta_r}( \| \textbf{x}^\alpha_i - \textbf{x}^\beta_j \| -r), \label{eq:pcf2}
\end{align}
where $\{\textbf{x}_i^\alpha\}_{i = 1,2,...,N_\alpha}$ are the points of type $\alpha$ in our data (i.e., the $(x,y)$-coordinates of light-blue points in Fig.~\ref{fig:pcf}(b)) and $\{\textbf{x}_i^\beta\}_{i = 1,2,...,N_\beta}$ are the points of type $\beta$ (i.e., the $(x,y)$-coordinates of navy points in Fig.~\ref{fig:pcf}(b)); the function $\mathds{1}_{\delta_r}(d) = 1$ if $0<d< \delta_r$ and $0$ otherwise; $A$ is the area of the domain of interest, which is the area of the image when working with image data; $A_r(\textbf{x})$ is the area of the annulus of inner radius $r$ and width $\delta_r$ centered at point $\textbf{x}$ and contained in our domain; and $\| \cdot \|$ is the Euclidean norm. Because I use periodic boundary conditions in the $x$- and $y$-directions in Fig.~\ref{fig:pcf}, $A_r(\textbf{x})$ simplifies to $\pi (r+\delta_r)^2 - \pi r^2$; however, in cases without periodic boundary conditions, more care is needed to compute the PCF \cite{Bull2020,PCFbook2}. (I have rewritten the equations in \cite{Bull2020} in my own notation in Eqns.~\eqref{eq:pcf1}--\eqref{eq:pcf2}.)

When the value of $\text{PCF}(r)$ or $\text{cross-PCF}(r)$ in Eqns.~\eqref{eq:pcf1}--\eqref{eq:pcf2} is greater than one, it means that points often appear at a distance of $r$ from one another in our data \cite{Bull2020,Bull2023wPCF}. On the other hand, a PCF of less than one indicates that points at this distance away from one another are less common than would be the case if the points were randomly dispersed. As I show in Fig.~\ref{fig:pcf}(c), the PCF for like points in Fig.~\ref{fig:pcf}(b) peaks first near the mean nearest-neighbor distance, and later dips around $r=300$. The cross-PCF, on the other hand, is less than one for low values of $r$, suggesting anti-correlation between unlike points locally, and reaches a maximum near $r=300$, which we can interpret as the characteristic width of the labyrinthine stripes in Fig.~\ref{fig:pcf}(b). Pair correlation functions are very flexible, and I show in Fig.~\ref{fig:pcf}(d)--(e) how they can be used to capture information about spot separation and size in patterns. Averaging PCFs is common when analyzing point-clouds generated from stochastic models or multiple experimental replicates (i.e., \cite{Agnew2014,Treloar2014spatial}).

The examples in Fig.~\ref{fig:pcf} illustrate how pair correlation functions can be applied to off-lattice data, but the PCF approach is also a natural means to quantify lattice point-cloud data, such as the positions of pixels in a segmented image like Fig.~\ref{fig:data}(a) or grid positions in an on-lattice agent-based model. For example, Binder and Simpson \cite{Binder2013} applied PCFs to non-overlapping, discrete data from an agent-based model and to pixel positions in binarized empirical images, and Dini et al. \cite{Dini2018} introduced a cross-PCF function for describing correlation and anti-correlation in the positions of agents of different types on a square lattice. Gavagnin et al. \cite{Gavagnin2018yates} developed PCFs for many tessellations, including hexagonal, triangular, and cuboidal lattices.

In some settings (i.e., \cite{Dini2016,Binder2015yeast,Gavagnin2018yates}), whether the data are on or off lattice, the Euclidean distance is not the most suitable for quantifying features of interest. With this motivation, a range of PCFs have been developed that consider distances between points in only one direction, such as in $x$, in $y$, or angularly \cite{Dini2016,Binder2015yeast}. For example, angular distances are appropriate for quantifying spreading and branching yeast patterns \cite{Binder2015yeast}. In other settings, the locations of points are constrained by external forces and obstacles; Johnston and Crampin \cite{Johnston2019pre} thus developed a PCF for environments that have obstacles, such is the case for pedestrians moving in rooms.

\subsection{Techniques from topological data analysis}
\label{sec:tda}

\begin{figure}[t!]
\includegraphics[width=0.7\textwidth]{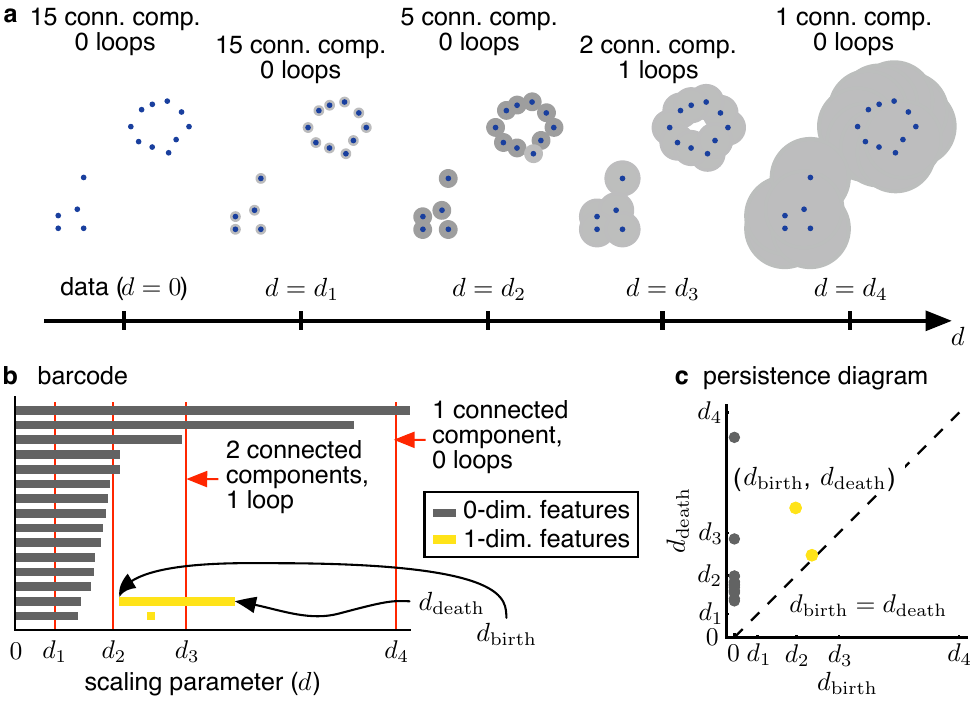}\centering
\caption{\label{fig:tda1}Introduction to persistent homology, a widely used tool in topological data analysis \cite{Topaz2015,Munch2020shape,Carlsson2009,Edelsbrunner2008,Otter2017mason}. (a) To understand the shape of the data in the left-most image, one can place balls around each point and increase the ball diameter $d$. The birth time of a topological feature is the value of the filtration parameter ($d$ here) at which the feature appears, and the death time is the first filtration-parameter value at which the feature no longer appears. (b) Barcodes and (c) persistence diagrams summarize how the number of connected components (dimension-$0$ topological features) and holes or loops (dimension-$1$ topological features) vary as the scaling parameter increases.}
\end{figure}

Persistent homology is a flexible tool in topological data analysis (TDA) that relies on ideas from algebraic topology to characterize shape \cite{Carlsson2009,Munch2020shape,Edelsbrunner2008,Topaz2015}. Persistent homology describes connected components (i.e., clusters), loops (i.e., holes), and higher-dimensional features in point-cloud data across scales \cite{Otter2017mason}. Notably, persistent homology can be applied to all of the images in Fig.~\ref{fig:summary} and provide meaningful quantitative information. It has been used to shed light on many biological patterns, including empirical and simulated data on fish-skin patterns \cite{McGuirl2020,Cleveland}, flocking and milling \cite{Topaz2015,Bhaskar2019}, vascular networks \cite{Nardini2021arxiv}, intracellular dynamics \cite{Ciocanel2021}, honeybee aggregation \cite{PelegTDA2023}, coral reefs \cite{Coral}, cancer tissue \cite{Lawson2019cancer}, and brain arteries \cite{Bendich2016}. In this subsection, I discuss persistent homology informally and give examples of how it can be applied to complex systems. See \cite{Carlsson2009,Edelsbrunner2008,Otter2017mason,Topaz2015,Ghrist2014} for technical details. I also recommend the recent review \cite{Munch2020shape} for an overview of TDA and its interactions with biology, and the article \cite{Otter2017mason} for guidance on computing persistent homology (along with a a discussion of various open-source software libraries involved).

\begin{figure}[t!]
\includegraphics[width=0.7\textwidth]{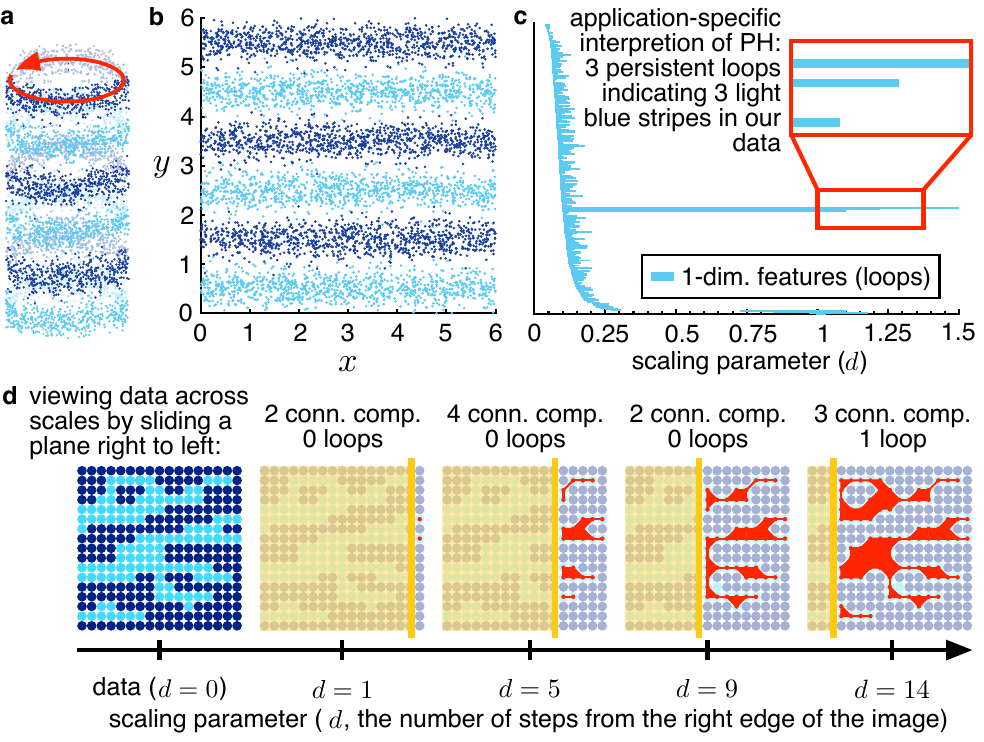}\centering
\caption{\label{fig:tda2}Examples of how persistent homology can be used to quantify stripe and branching patterns. (a) After computing persistent homology, the next step is interpreting barcodes and persistence diagrams. As an example, McGuirl et al. \cite{McGuirl2020} showed how to apply persistent homology to stripe-pattern data with periodic boundary conditions in one direction. In this case, the stripes wrap around a cylinder. (b) I unwrap the cylinder, indicating three light-blue and three navy stripes, and (c) show the associated barcode for dimension-$1$ topological features, computed using the positions of light-blue points. The three long bars---i.e., three highly persistent features---correspond to three light-blue stripes \cite{McGuirl2020,Cleveland}. (d) There are many other choices of filtration in persistent homology, and this choice is application-dependent \cite{Munch2020shape}. In (c) and the example in Fig.~\ref{fig:tda1}, for example, I compute persistent homology by placing balls around each point in the point cloud and increasing the ball size. In the work \cite{Nardini2021arxiv,NardiniRetinal} of Nardini et al. on branching patterns, a sweeping-plane filtration is more natural. This can be thought of as slowly uncovering an image by sweeping a plane across it, and counting the number of connected components and holes that emerge during this process.}
\end{figure}

In order to compute persistent homology, one needs to filter through data using some scaling parameter or threshold. I illustrate a common approach to doing this that involves building a Vietoris--Rips complex in Fig.~\ref{fig:tda1}(a). Suppose I have $15$ points in my pattern data, and these points, which I show in the left panel of Fig.~\ref{fig:tda1}(a), are the $(x,y)$-coordinates of $15$ people in a room. How many groups are in the room? Perhaps you see a ring of people talking, one cluster, and a lone individual, or you may judge this pattern to contain two groups only. Persistent homology provides a means of quantifying the shape of the positions of people in the room without the need for this judgment call. In order to build a Vietoris--Rips complex, we place a ball centered at each point and slowly increase the ball diameter $d$. As the scaling parameter $d$ grows, we keep track of the appearance and disappearance of connected components (dimension-$0$ features) and holes (dimension-$1$ features). The result is often visualized using barcodes and persistence diagrams; see Fig.~\ref{fig:tda1}(b) and Fig.~\ref{fig:tda1}(c), respectively. For example, in order to plot a barcode describing the $1$-dimensional topological features in data, one draws a horizontal bar for each of these features, where the bar starts at the value $d_\text{birth}$ of the scaling parameter at which the feature appears (the feature's ``birth time") and ends at the value $d_\text{death}$ of the scaling parameter at which the feature disappears (the feature's ``death time") \cite{Topaz2015}.

Once data have been summarized as in Fig.~\ref{fig:tda1}(b), the next step is to interpret and use the resulting barcode \cite{Otter2017mason}. For example, McGuirl et al. \cite{McGuirl2020} interpreted the number of highly persistent dimension-$0$ and dimension-$1$ features (i.e., long bars in barcodes) as the number of spots and stripes, respectively, in simulated zebrafish-skin patterns. I highlight how persistent homology can be used to count the number of stripes on a domain with periodic boundary conditions in $x$ in Fig.~\ref{fig:tda2}(a)--(c). (Importantly, this process involves the choice of hyper-parameters \cite{Cleveland}.) In other cases, the shape of data is not as clear as in Fig.~\ref{fig:tda2}(b), and persistence landscapes can be useful in this setting \cite{Bubenik2015,CruzArxiv}. Briefly, a persistence landscape is a decreasing sequence of functions that is defined using the information in persistence diagrams, and persistence landscapes lend themselves to averaging results across patterns and work well with machine learning \cite{Bubenik2015,CruzArxiv}. Persistence landscapes are not the only means of making the output of persistent-homology computations appropriate for machine learning, and I suggest the survey \cite{Ali2023} for more information.

Combining persistent homology with machine learning is a very active area, and references \cite{Bonilla2020,Bhaskar2019,Bhaskar2021,Bhaskar2023,CruzArxiv,Nardini2021arxiv} are some examples of this process across various biological systems. For instance, Bonilla et al. \cite{Bonilla2020} applied persistent homology to simulated and real data on collective cell motion for moving interfaces, and they used unsupervised learning to cluster different types of patterns. Working with simulated and empirical patterns that exhibit branching and clustering, Bhaskar et al. \cite{Bhaskar2021} partitioned their data by applying hierarchical clustering based on the pairwise Wasserstein distances between persistence diagrams. As another example, Nardini et al. \cite{NardiniRetinal} used persistent homology together with machine learning to characterize and classify segmented images of retinal vasculature.

Just like specific order parameters (see \S\ref{sec:order}) and pair-correlation functions (see \S\ref{sec:pair}) are more or less appropriate for various patterns, the method for filtering through data in Fig.~\ref{fig:tda1}(a) is not the most meaningful approach to persistent homology for all patterns. In fact, there are many ways of filtering through data across scales \cite{Munch2020shape}. As an example that is useful for quantifying branching, I highlight the sweeping-plane filtration in Fig.~\ref{fig:tda2}(d) \cite{Nardini2021arxiv}. Nardini et al. \cite{Nardini2021arxiv} applied this filtration, among others, to branching patterns generated by a model \cite{Anderson1998} of angiogenesis. In this case, rather than varying the size of a ball around points, one sweeps through an image, uncovering more of the picture and tracking the topological features that emerge during this process. Persistent homology can also be applied to grayscale images, even without thresholding as in Fig.~\ref{fig:data}(a), by filtering through pixels by intensity \cite{Munch2020shape}.

\section{Discussion and outlook}
\label{sec:discussion}

To conclude, I discuss three directions related to quantifying patterns that I see as particularly exciting areas of current and future research. First, while I presented approaches to quantifying complex systems in separate subsections above, many recent studies \cite{Nardini2021arxiv,Bull2020,Bull2024preprint,Topaz2015,Bhaskar2019} have brought multiple quantitative perspectives to bear on their data. For example, Nardini et al. \cite{Nardini2021arxiv} quantified angiogenesis used two different persistent-homology filtrations, as well as direct measurements. In a series of papers \cite{Bull2020,Bull2024preprint}, Bull and collaborators combined a wide range of methods---including pair-correlation functions, other approaches from spatial statistics, and techniques from topological data analysis---to better understand cell organization and cancer data. And Topaz and collaborators \cite{Topaz2015,Ulmer2019topaz,Bhaskar2019} investigated persistent homology and order parameters for aggregation. Studies that bring methods together allow researchers to take a broader perspective on their data, identify the relationship between approaches, and determine how the choice of method may impact conclusions. 

Second, investigating the role of choices and hyper-parameters in quantifying data is an important direction for future work, as all of the quantification approaches in \S\ref{sec:methods} involve choices. For example, in addition to choosing what pair correlation function to apply to data, one must also choose the annulus width $\delta_r$. Binder and Simpson \cite{Binder2015} discuss the impact of selecting a bandwidth when working with pair-correlation functions and on-lattice point clouds, for instance. In the case of persistent homology in Fig.~\ref{fig:tda1}(c), interpreting barcodes often requires the choice of several hyper-parameters, as Cleveland et al. \cite{Cleveland} note. The wealth of quantitative methods available to describe pattern data opens up the door to understanding variability across stochastic simulations \cite{McGuirl2020} and biological images, and it is important to understand how choices in the quantification pipeline affect results. 

Lastly, there are many exciting directions at the intersection of data-driven modeling and pattern quantification. For example, Johnston et al. \cite{Johnston2014} estimated parameters in an agent-based model of collective cell spreading, combining pair correlation functions with cell counts to quantify their data. In the case of branching patterns in kidney tissue, Lambert et. al. \cite{Lambert2018aabc} used direct measurements of area and branch points in time as their summary statistic before performing Bayesian-style inference in an agent-based model. From the topological-data-analysis perspective, Thorne et al. \cite{Thorne} combined persistent homology with approximate Bayesian methods to infer parameters in a model of angiogenesis. Recently equation-learning and model-selection approaches (e.g., \cite{Mangan2017,Brunton2016,Nardini2021,Kemeth2022,Lagergren2020,Gonzalez1998,Raissi}) have emerged as active research directions. In this vein, Lagergren et al. \cite{Lagergren2020} used neural-network approaches to learn partial differential equations describing cell migration in scratch-assay experiments, evaluating the match via one-dimensional cell profiles. And Liu et al. \cite{Yue2024} highlighted how choices in quantifying noisy data affect parameter inference and identifiability in the case of continuum models of cell spreading. Building inference and data-driven methods for complex models of spatial pattern dynamics are exciting directions for future work, and quantifying qualitative data is a key part of this process.\\

\noindent \textbf{Acknowledgments:}
A.V.\ was a programme participant in the ``Mathematics of movement: an interdisciplinary approach to mutual challenges in animal ecology and cell biology" programme at the Isaac Newton Institute for Mathematical Sciences, Cambridge, during fall 2023. She gratefully acknowledges the support, hospitality, and excellent environment of the Isaac Newton Institute, as well as a travel grant from the Association for Women in Mathematics. A.V.\ also thanks the Mathematics of movement organizers for organizing the Isaac Newton Institute programme where work on this paper was undertaken. This work was supported by EPSRC grant EP/R014604/1.\\

\noindent \textbf{Competing Interests:} {A.V.\ has no conflicts of interest to declare that are relevant to this chapter.}\\

\noindent \textbf{Ethics Approval:} {No ethical approval is necessary for the subject of this chapter.}

{}

\end{document}